\pdfoutput=1

\documentclass[3p,times]{elsarticle}

\usepackage{ecrc}

\usepackage[breaklinks]{hyperref}
\usepackage[utf8x]{inputenc}
\usepackage{subfig}
\usepackage{amsmath}
\usepackage{graphicx}

\volume{00}

\firstpage{1}

\journalname{Nuclear Physics A}

\runauth{Ye-Yin Zhao, Ming-Mei Xu, Heng-Ying Zhang and Yuan-Fang Wu}

\jid{nupha}

\jnltitlelogo{Nuclear Physics A}

\usepackage{amssymb}

\biboptions{comma,square,sort&compress}

\usepackage[figuresright]{rotating}

\def\eq#1{{Eq.~(\ref{#1})}}

\def\bp{\bold{p}}
\def\bq{\bold{q}}
\def\bk{\bold{k}}
\def\br{\bold{r}}
\def\brt{\bold{r_{\perp}}}

\begin{document}



\hypersetup{pdfauthor={Ye-Yin Zhao, Ming-Mei Xu, Heng-Ying Zhang and Yuan-Fang Wu},pdftitle={Two-gluon rapidity correlations of strong colour field in $pp$, $pA$ and $AA$ collisions}}



\dochead{}


\author{Ye-Yin Zhao, Ming-Mei Xu, Heng-Ying Zhang and Yuan-Fang Wu}
\address{
Key Laboratory of Quark and Lepton Physics (MOE) and Institute
of Particle Physics, Central China Normal University, Wuhan 430079, China
}

\title{
Two-gluon rapidity correlations of strong colour field in $pp$, $pA$ and $AA$ collisions
}

\begin{abstract}
Using the CGC formalism, we calculate the two-gluon rapidity correlations of strong colour fields in $pp$, $pA$ and $AA$ collisions, respectively. If one trigger gluon is fixed at central rapidity, a ridge-like correlation pattern is obtained in symmetry $pp$ and $AA$ collisions, and a huge bump-like correlation pattern is presented in asymmetry $pA$ collisions. It is demonstrated that the ridge-like correlation patterns are caused by the stronger correlation with the gluon of colour source. The transverse momentum and incident energy dependence of the ridge-like correlation pattern is also systematically studied. The ridge is more likely observed at higher incident energy and lower transverse momentum of trigger gluon.
\end{abstract}
\maketitle

\section{Introduction}\label{sec:introduction}
	
The Color Glass Condensate (CGC) effective field theory (EFT) provides an identical description
for the initial state of high energy scattering in quantum chromodynamics(QCD) framework~\cite{Gribov:1983PR100,Weigert:2005ppnp,Iancu:2003cgc,Gelis:2010annu}. Assuming that final state interactions can either be computed or that they are in some circumstances small corrections~\cite{Larry-WWND2013}, it can be applied to study the multiparticle production at a number of existing and upcoming high energy colliding facilities.

The Color Glass Condensate is weakly coupled, but strongly interacting highly coherent gluonic matter in the wave function of a nucleus. The effective degrees of freedom in this framework are color sources $\rho$ at large $x$ and gauge fields $\mathcal{A}_{\mu}$ at small $x$. Given an initial distribution of sources at an initial scale $x_0$, its dynamical evolution are captured by the Balitsky-Kovchegov (rcBK) equation. Gluon production in this stage is saturated with typical momenta peaked at a characteristic value $k_{\perp}\propto Q_s$, and localized in typical size $r_{\perp s}\sim 1/Q_s$. The occupation number of gluon $1/\alpha_s$ is large. So it is strongly interacting and highly coherent colour field. 

In nuclear collisions, CGC dynamics produces “Glasma” field configurations at early times: strong longitudinal chromo-electric and chromo-magnetic fields color screened on transverse distance scales $r_{\perp s}\sim 1/Q_s$. Where $Q_s$ grows with the energy and centrality of the collision~\cite{Kowalski:2006PRD74,Levin:2010PRD82,Dusling:1509.04410}. Multi-particles are produced in this approximately boost invariant Glasma flux tube, which generates long range longitudinal correlations.

Using the CGC EFT and combining the fragmentation models of hadronization, the multiplicity distributions in $pp$ and $pA$ collisions at RHIC through LHC energies are successfully described~\cite{Dumitru:2006npa,Dumitru:2006npa1,Goncalves:2006plb,Tuchin:2008npa,Albacete:2010plb,Rezaeian:2010prd,Fujii:2011jpg,Albacete:2012npa}. The transverse momentum distributions of charged particles for $pp$ collisions at LHC energies are also described~\cite{Albacete:2012npa}. The transverse momentum broadening predicted by the CGC as a function of multiplicity is clearly seen in the data~\cite{McLerran:2010appb}.  

The data of ridge-like dihadron correlations at RHIC and LHC~\cite{STAR:2005PRL95,PHENIX:2008PRC78,PHOBOS:2010PRL104,CMS:2010JHEP,CMS:2012PLB,ALICE:2013PLB,CMS:2013PLB724} can be qualitatively understood by some of the phenomenological models with the early-time dynamics determined by CGC~\cite{Dumitru:2008NPA810,Dusling:2010NPA836,Dumitru:2011PLB697,
Gelis:2009PRD79,Dumitru:2010PRD81,Dusling:2012PRL108,Dusling:2013PRD87,Dusling:2013PRD87-1,Dusling:2013PRD87-2}. However, although the computation of CGC show qualitatively the trend of ridge-like correlation, it is still not clear what should be responsible for the ridge-like correlations in the initial state~\cite{Larry-WWND2013,Dusling:1509.07939}. Is it the maximal value of the topological charge density when two trigger gluon are collimation~\cite{Dusling:2013PRD87,Dusling:2013PRD87-1,Dusling:2013PRD87-2}, the initial state quantum mechanical effects, and/or the fluctuations of a colour flux tube in the longitudinal space~\cite{Larry-WWND2013}?

In order to see why ridge-like correlation in initial state, we focus our calculations on the gluon level alone and use only the formalism of CGC EFT~\cite{Dumitru:2008NPA810,Dusling:2010NPA836,Dumitru:2011PLB697,Gelis:2009PRD79,Dusling:2012PRL108}. In the framework of CGC, the mechanism of gluon production varies with $x$ degree of freedom. It should be helpful if we show the contributions of the various $x$ degrees of freedom in the two-gluon rapidity correlation pattern. 

On the other hand, the CGC EFT provides an identical description for proton and nucleus. The origin of the ridge-like correlation should be independent of colliding particles. So we present our calculations for as many as possible processes, e.g., $pp$, $pA$ and $AA$ collisions. They will enhance our understanding on the origin of ridge-like correlation in the initial state.
 
In this paper, we first introduce in section II the calculations of two-gluon rapidity correlations in the framework of CGC EFT. The unintegrated gluon distribution (uGD), and its characteristic values for proton and nucleus are presented in section III. 

In section IV, the transverse momentum dependence of two-gluon rapidity correlations for $pp$, $pA$ and $AA$ collisions are presented. They show that the correlations have the maximum when the transverse momentum of two gluon are both close to the sum of saturation momenta of two colliding particles. Their features in different processes reveal the properties of the uGD of associated colliding particles. 

In section V, the two-gluon rapidity correlations as the function of rapidity gap in minimum bias samples of $pp$, $pA$ and $AA$ collisions are presented. If one trigger gluon is fixed at central rapidity, a ridge-like correlation pattern is obtained in symmetry $pp$ and $AA$ collisions, and a huge bump-like correlation pattern is presented in asymmetry $pA$ collisions. It is demonstrated that the ridge-like correlation patterns are caused by the stronger correlation with the gluon of colour source. The transverse momentum and incident energy dependence of ridge-like correlation pattern is systematically studied.  

Finally, a brief summary is given in section VI.
	
\section{Two-gluon rapidity correlations in the framework of CGC}\label{sec:applications}
The correlation of two gluon with transverse momentum $\bp_\perp$ and $\bq_\perp$ and longitudinal rapidity $y_p$ and $y_q$ is defined as~\cite{Dumitru:2011PLB697}, 
\begin{equation}\label{eq:correlation}
C(\bp_{\perp},y_p;\bq_{\perp},y_q)=
\frac{\frac{dN_2}{d^2\bp_{\perp}dy_pd^2\bq_{\perp}dy_q}}
{\frac{dN_1}{d^2\bp_{\perp}dy_p}\frac{dN_1}{d^2\bq_{\perp}dy_q}}-1.
\end{equation}
In the framework of CGC EFT, the two- and single-gluon distributions for $pp$, $AA$ and $pA$ collisions at the leading logarithmic accuracy in $x$ can be identically expressed by~\cite{Dusling:2010NPA836,Kovchegov:1998NPB529,Fujii:2006NPA780},
\begin{equation}\label{eq:two-gluon}
\frac{dN_2}{d^2\bp_{\perp}dy_pd^2\bq_{\perp}dy_q}=
\frac{\alpha^2_s}{16\pi^{10}}\frac{N^2_c}{(N^2_c-1)^3p^2_{\perp}q^2_{\perp}}
\int_0^{\infty}{d^2\bk_{\perp}}(D_1+D_2),
\end{equation}  
and 
\begin{equation}\label{eq:single-gluon}
	\frac{dN_1}{d^2\bp_{\perp}dy_p}=\frac{\alpha_sN_c}
	{\pi^4(N^2_c-1)}\frac{1}{p^2_{\perp}}
	\int\frac{d^2\bk_{\perp}}
	{(2\pi)^2}\Phi_A(y_p,\bk_{\perp})\Phi_B(y_p,\bp_{\perp}-\bk_{\perp}).
\end{equation}
Where $N_c=3$ is the number of colors, and
\begin{eqnarray}\label{eq:d1d2}
	D_1=\Phi^2_A(y_p,\bk_{\perp})\Phi_B(y_p,\bp_{\perp}-\bk_{\perp})D_B,\\
	D_2=\Phi^2_B(y_q,\bk_{\perp})\Phi_A(y_p,\bp_{\perp}-\bk_{\perp})D_A,
\end{eqnarray}
with $D_{A(B)}=\Phi_{A(B)}(y_q,\bq_{\perp}+\bk_{\perp})+\Phi_{A(B)}(y_q,\bq_{\perp}-\bk_{\perp})$.


$\Phi$ is \textit{unintegrated gluon distribution} (uGD), i.e., gluon density at an unit transverse area per rapidity of proton, or nucleus $A$, or $B$. It is related to dipole forward scattering amplitude, and can be expressed in large $N_c$ limit as~\cite{Dusling:2010NPA836},
\begin{equation}\label{eq:uGD}
\begin{split}
	\Phi_{A(B)}(x,\bk_{\perp})&=\frac{N_ck^2_{\perp}}{4\alpha_s}\int d^2\br_{\perp}e^{i\bk_{\perp}\cdot \br_{\perp}}\left[1-\mathcal{N}_{\text{ad.}}(\br_{\perp},Y)\right]\\
	&=\frac{\pi N_c k^2_{\perp}}{2\alpha_s}\int dr_{\perp}r_{\perp}J_0(k_{\perp}r_{\perp})\left[1-\mathcal{N}(r_{\perp},Y)\right]^2.
\end{split}
\end{equation}
where $J_0$ is Bessel function, $Y=\ln\frac{x_0}{x}=\ln x_0-\ln x$ is the rapidity interval of produced gluon, $x$ is the fraction of longitudinal momentum of produced gluon in the infinite momentum frame (IMF), and $x_0$ is starting value of small-$x$ dynamic evolution and often sets $x_0=0.01$.

$\mathcal{N}$ is quark-antiquark dipole forward amplitude in fundamental representation. In large-$N_c$ limit, its quantum evolution towards small-$x$ is described by running coupling Balitsky-Kovchegov(rcBK) equation~\cite{Balitsky:1996NPB463,Kovchegov:1999PRD60,Albacete:2007PRD75}. It incorporates all multi-scattering between a projectile color dipole and the target, and the small-$x$ dynamics. i.e.,
\begin{equation}\label{eq:rcbk}
	\frac{\partial\mathcal{N}(\brt,x)}{\partial\ln(x_0/x)}=
	\int d^2\brt_{1}K^{\text{run}}(\brt,\brt_1,\brt_2)
	\Big[\mathcal{N}(\brt_1,x)+\mathcal{N}(\brt_2,x)-\mathcal{N}(\brt,x)-\mathcal{N}(\brt_1,x)\mathcal{N}(\brt_2,x)\Big].
\end{equation}
where $\brt=\brt_1+\brt_2$ is the transverse dipole size. $K^{\text{run}}$ is running coupling kernel. It slows the evolution significantly, and is necessary in describing Deep Inelastic Scattering (DIS) and hadron scattering data. According to Balitsky's prescription\cite{Balitsky:2007PRD75,Albacete:2011EPJC71}, it is as follows:
\begin{equation}
	K^{\text{run}}(\brt,\brt_1,\brt_2)=\frac{N_c\alpha_s(\brt^2)}{2\pi^2}\Bigg[\frac{1}{\brt^2_1}\Bigg(\frac{\alpha_s(\brt^2_1)}{\alpha_s(\brt^2_2)}-1\Bigg)+\frac{\brt^2}{\brt^2_1 \brt^2_2}+\frac{1}{\brt^2_2}\Bigg(\frac{\alpha_s(\brt^2_2)}{\alpha_s(\brt^2_1)}-1\Bigg)\Bigg].
\end{equation}
We use its one-loop approximation in our calculations. 
	
Usually, the amplitude $\mathcal{N}$ is obtained by solving rcBK evolution equation at a given initial condition. In our calculations, we use the parametrization of the Albacete-Armesto-Milhano-Quiroqa-Salgodo (AAMQS) model~\cite{Albacete:2011EPJC71},
\begin{equation}\label{eq:aamqs}
	\mathcal{N}_{\text{AAMQS}}(r,x_0=0.01)=1-\exp\Big[-\frac{1}{4}\big(r^2Q^2_s(x_0)\big)^{\gamma}\ln\bigg(e+\frac{1}{r\Lambda}\bigg)\Big],
\end{equation}
where $\Lambda=0.241\ \mathrm{GeV}$ is the infrared scale. The anomalous dimension $\gamma$ and the initial saturation scale $Q_s(x_0)$ are free parameters. The best fit of data requires that $\gamma > 1$. $\gamma=1$ corresponds to the case of McLerran-Venugopalan (MV) model~\cite{MV}. In our calculations, we adopt $\gamma=1.119$ for both proton and nucleus~\cite{Albacete:2011EPJC71}.

The initial saturation scale $Q_s(x_0)$ changes only with colliding particles and 
centrality~\cite{Kowalski:2006PRD74,Levin:2010PRD82}. For proton, the initial condition $Q_{s0p}$ is determined by a global fit to the proton structure function $F_2$ in DIS, and to the single inclusive hadron data of $pp$ collisions at the RHIC and the LHC. These require $Q^2_{s0p}=0.168\mathrm{GeV}^2$ for minimum bias samples. It also provides a good description for charged hadron transverse momentum distributions in $pp$ collisions at the center-of-mass energy per nucleon pair $\sqrt{s_{NN}}=7\ \mathrm{TeV}$~\cite{Albacete:2010xrXiv5161,Dumitru:2012PTPS193}. 

For a nucleus, the initial saturation scale $Q_{s0A}$ is often determined by $Q^2_{s0A}=c(b)A^{1/3}Q^2_{s0p}$~\cite{Dusling:2010NPA836}. In our calculations, we set it three times that of the proton, i.e., $Q^2_{s0A}=0.504\ \mathrm{GeV}^2$ for minimum bias samples.

It should be noticed that the values of $Q_{s0}$ for proton and nucleus are not invariable. 
The relations between initial saturation scale $Q_{s0}$ and hadron productions depend on the impact parameter, and has been systematically studied~\cite{Kowalski:2006PRD74,Levin:2010PRD82}.
	
\section{The uGD in the framework of CGC}
	
From \eq{eq:two-gluon} and \eq{eq:single-gluon}, we can see that two- and single-gluon distributions are related to the convolution of four and two uGDs of projectile and target particles, respectively. Therefore, uGD plays a key role in the two-gluon rapidity correlation. It is a universal quantity in describing high energy evolution towards small-$x$ dynamics, and can be obtained by solving rcBK equation. The only parameter is the initial saturation scale $Q_{s0}$ in a given collision. 

To see its general feature, as an example, we present the uGD of a proton at the plane of rapidity $Y$ and transverse momentum $k^2_{\perp}$ in Fig.~1(a). When the uGD increases from zero to above 3, the colour of the plot changes from dark blue to light yellow as indicated in the legend. It has a bell-shaped structure. There is a peak in both $Y$ and $k^2_{\perp}$ directions. 

To see how the peak positions of uGD changes with $Y$ and $k^2_{\perp}$, the profile of Fig.~1(a) at given $Y=0$ (red line), 2(blue dash line), 5 (purple dash line), and 10 (grey dash line) are presented in Fig.~1(b), respectively. It shows the distribution of gluon at a given $Y$, or $x$. The peak position of uGD corresponds to the saturation scale $Q_s$, and this value is also the typical transverse momentum of the gluon in the hadron wave function. It also shows that with the increase of $Y$, the peak position moves to larger $k^2_{\perp}$, and the whole uGD becomes lower and broader. 

\begin{figure}[htb]
		\includegraphics[width=\linewidth]{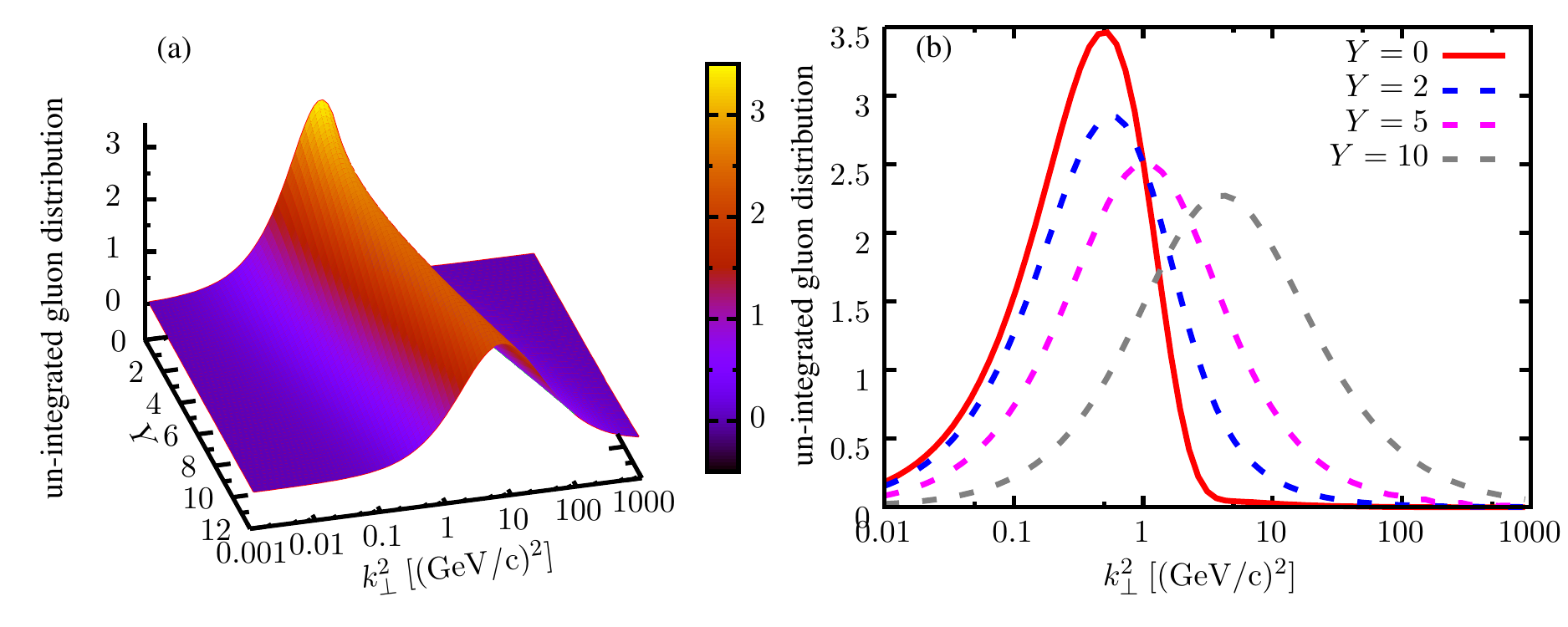}
		\caption{(a) The uGD of proton in $Y$ and $k^2_{\perp}$ plane, where the value of uGD is indicated in the legend by colour. (b) The uGD of proton at given $Y=0$ (red line), 2(blue dash line), 5(purple dash line), 10 (grey dash line).}
		\label{fig:ugdall}
\end{figure}

The saturation scale $Q_s$ characterizes the peak position of uGD, and changes with rapidity. For proton and nucleus, we present their rapidity dependence of saturated momentum scale $Q_s$ in Fig.~2(a) by red and blue solid lines, respectively. Where the initial saturated momentum for proton and nucleus is $Q^2_{s0}=0.168\ \mathrm{GeV}^2$ and $0.504\ \mathrm{GeV}^2$. It shows that the saturation scale $Q_s$ increases slowly with rapidity in both proton and nucleus at small rapidity.  At a given rapidity, the saturated momentum scale $Q_s$ of the nucleus is larger than that of a proton.
	
\begin{figure}[htb]
		\includegraphics[width=\linewidth]{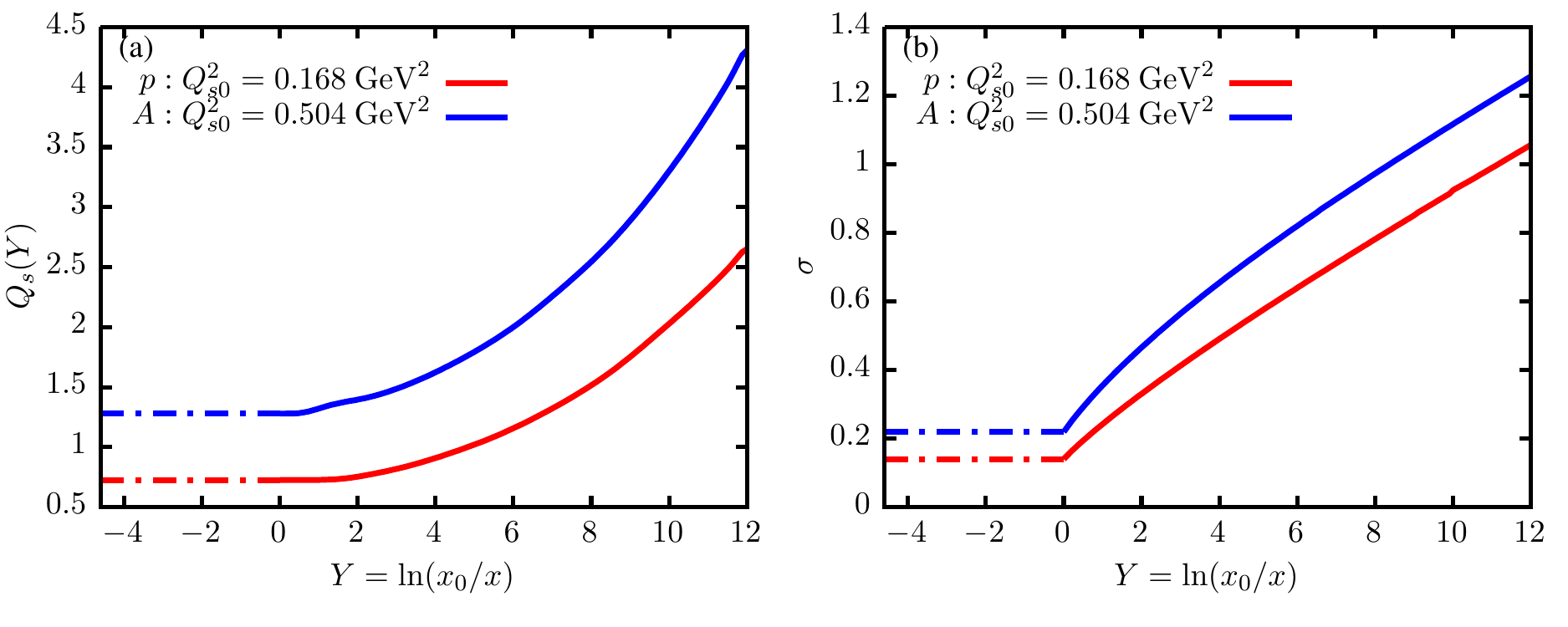}
		\caption{Rapidity dependence of the saturation scale $Q_s$ (a) and the widths (b) of uGD of proton (red dash-point and solid lines) and nucleus (blue dash-point and solid lines). Where the initial saturation scales for proton and nucleus are $Q^2_{s0}=0.168\ \mathrm{GeV}^2$ and $0.504\ \mathrm{GeV}^2$, respectively. }
		\label{fig:saturation}
\end{figure}	

Meanwhile, we present the widths ($\sigma$) of uGD of proton (red line) and nucleus (blue line) in Fig.~2(b). It shows that the width of uGD also increases with rapidity in both proton and nucleus. At a given rapidity, the width of uGD of a nucleus is wider than that of a proton. 

So for proton and nucleus, the saturation scale $Q_s$ (the widths $\sigma$ of uGD) has similar rapidity dependence. At a given rapidity, the difference of $Q_s$ ($\sigma$) between proton and nucleus is determined by the initial saturation scale $Q_{s0}$. This shows that the description for proton and nucleus in the framework of CGC EFT is indistinguishable, and the difference is only the initial condition. 

\section{Transverse momentum dependence of two-gluon rapidity correlations}
It is known that the correlation is the maximum when the transverse momentum of two selected gluon is collimated, i.e., $\Delta\phi=\phi_q-\phi_p=0,\pi$~\cite{Dumitru:2011PLB697,Dusling:2012PRL108}. But at what values of transverse momentum they have the strongest correlation is not obvious. This in fact requires that the kernel of integration of Eq.\eqref{eq:two-gluon} is the largest, i.e., the maximum of $D_1$, and/or $D_2$. As we known, the maximum of $\Phi_A(y_p, \bk_{\perp})$ appears at $\vert\bk_{\perp}\vert\sim Q_{sA}$. If we require in the same time that $\Phi_B(y_p, \bp_{\perp}-\bk_{\perp})$ and $\Phi_B(y_p, \bq_{\perp}-\bk_{\perp})$ are both maximum, then $\vert\bp_{\perp}\vert=\vert\bq_{\perp}\vert \sim Q_{sA}+Q_{sB}$. 

To demonstrate this simple estimation, we present in Fig.~3 the transverse momentum dependence of two-gluon rapidity correlations at a given rapidity gap for $\Delta\phi=0$ and $p_\perp=q_\perp$. Where Fig.~3(a), 3(b) and 3(c) are the results of minimum bias samples of $pp$, $AA$ and $pA$ collisions at $\sqrt s=7$ TeV, respectively. The rapidity of one gluon is fixed at $y_p=0$, and the rapidity gaps are chosen at $\Delta y=0$ (red line), 1 (blue line), 2 (purple line), 3 (green line), and 4 (black line), respectively. 

\begin{figure*} 
	\includegraphics[width=\linewidth]{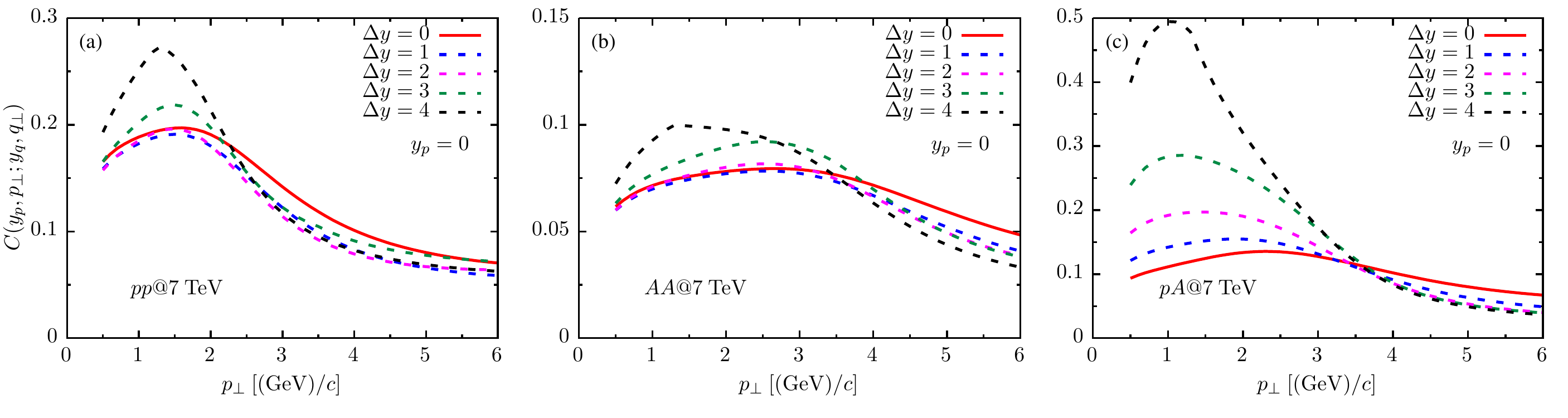}
	\caption{Transverse momentum dependence of $C(y_p,p_\perp;y_q,q_\perp)$ as the function of rapidity gap for minimum bias samples of $pp$ (a), $AA$ (b), and $pA$ (c) collisions at $\sqrt s= 7$ TeV. Where, $\Delta\phi=\phi_q-\phi_p=0$, $y_p=0$, $p_\perp=q_\perp$, and rapidity gaps $\Delta y=y_q-y_p=0$ (red line), 1 (blue line), 2 (purple line), 3 (green line), and 4 (black line).}
	\label{fig:fig211}
\end{figure*}

From the above section, we know that the saturated momentum of uGD changes with the rapidity interval of the produced gluon in the infinite momentum frame (IMF), i.e., $Y=\ln \frac{x_0}{x}$. Where $Y$ can be translated to the rapidity of two produced gluon in the center of mass frame (CMF)~\cite{Iancu:2003cgc}, $y_p$, and $y_q$, i.e.,
\begin{eqnarray}\label{eq:relations between Y and y}
Y_{pA(B)}=\ln\frac{\sqrt s}{p_{\perp}}+\ln x_0\pm y_p,\nonumber \\ Y_{qB(A)}=\ln\frac{\sqrt s}{q_{\perp}}+\ln x_0\mp y_q.
\end{eqnarray}
where, 
\begin{eqnarray}\label{eq:rapidity in cms}
	x_{Ap}=\frac{p_{\perp}}{\sqrt{s}}e^{-y_p},\qquad x_{Bp}=\frac{p_{\perp}}{\sqrt{s}}e^{+y_p}, \nonumber\\
	x_{Aq}=\frac{q_{\perp}}{\sqrt{s}}e^{+y_q},\qquad x_{Bq}=\frac{q_{\perp}}{\sqrt{s}}e^{-y_q},
\end{eqnarray}
In the central rapidity region, i.e., $y_p=y_q=0$, for $\sqrt s=7$ TeV and $p_{\perp}, q_{\perp} \sim 1$ GeV, corresponding $Y \sim 4.25$. From Fig.~2(a), when $Y\sim 4$, the $Q_s$ for the proton and nucleus are around 0.8 GeV and 1.4 GeV, respectively. Both of them decrease very slightly and slowly with the decrease of $Y$. 

For symmetry $pp$ collisions in Fig.~3(a), the maximum correlation for $\Delta y=0$ just appears at $p_{\perp}\sim 2Q_{sp}=1.6$ GeV. With increase of the rapidity gap, i.e., increase of $y_q$, the corresponding $Y_{qB(A)}$ decreases, and so does the $Q_{sp}$. Therefore, the peak of the correlation should move slightly and slowly toward small transverse momentum side, the same as $Q_{sp}$ does. This is just shown by blue, purple, green, and black lines in Fig.~3(a). 

The same features can also be seen in Fig.~3(b) for symmetry $AA$ collisions. While, the  difference is that the peak position for $\Delta y=0$ appears at larger $p_{\perp}\sim 2Q_{sA}=2.8$ GeV. 

The Fig.~3(a) and 3(b) also show that the correlations in symmetric $pp$ collisions are stronger than those in $AA$ collisions. This is simply due to the coupling constant $\alpha_{sp}$ in $pp$ are larger than that $\alpha_{sA}$ in $AA$ collisions. It will be more obviously shown in Fig.~4. 

The correlations of $pp$ collisions in Fig.~3(a) are spread in a narrower transverse momentum region in comparison to those of $AA$ collisions in Fig.~3(b). This can be directly understood by the width of uGD of proton and nucleus, as showed in Fig~2(b). Where, at a given rapidity, the width of uGD of a proton is narrower than that of the nucleus. This transforms to the two-gluon rapidity correlations. So the transverse momentum dependence of two-gluon rapidity correlations for $pp$ collisions is narrower than those for $AA$ collisions.

For asymmetric $pA$ collisions as showed in Fig.~3(c), the maximum for $\Delta y=0$ is also nearby $p_{\perp}=Q_{sp}+Q_{sA}\sim 2.2$ GeV, as expected. On the other hand, the correlations are more sensitive to the change of the rapidity gap, in contrast to those in symmetric $pp$ and $AA$ collisions. This is because the colliding proton and nucleus are in two opposite sides of rapidity, and so they produce gluon. The two-gluon correlation of proton is stronger than that of nucleus. Therefore, it changes constantly along the whole covered rapidity region in $pA$ collisions, in contrast to symmetric collisions. It will be fully displayed in the next section. 
   
Therefore, the strength, the peak position, and the width of the transverse momentum dependence of two-gluon rapidity correlations in these three processes are closely related to the distribution of uGD of colliding particles. 

For even larger rapidity gap, such as $\Delta y=3$, and 4, the correlations in three processes increase fast and separate from those curves at small rapidity gaps, and the peak positions move toward small $p_{\perp}$ side. These characters relate to the appearance of the ridge of large rapidity gap, and will be discussed in the next section. 
	
\section{Two-gluon rapidity correlations}
In the formalism of CGC, the mechanism of gluon production in the various regions of phase space are different~\cite{Gribov:1983PR100,Weigert:2005ppnp,Iancu:2003cgc,Gelis:2010annu}. The smaller the $x$ of gluon, the later the gluon evolves. This implies the larger saturated scale, the smaller coupling constant and weaker correlation. The gluon at the central rapidity region reflects the properties of the small-$x$ ($x<10^{-3}$) degrees of freedom. Where the quantum evolutions are essential. While the gluon at middle rapidity region presents the features of moderate-$x$ ($10^{-3}< x < 10^{-2}$) degree of freedom. Where the quantum effects become weak, and the contributions of colour source fields have to be taken into account. 

The gluon at large rapidity region shows the characters of larger-$x$ ($x>10^{-2}$) degree of freedom. Where the quantum evolution can be neglected in the first approximation, and the colour source fields is described by the MV model with Gaussian weight function. In the case, a phenomenological extrapolation for the uGD is used~\cite{Fujii:2006NPA780}, i.e.,

\begin{equation}\label{eq:uGD-large-x}
\begin{split}
	\Phi_{A(B)}(x,\bk_{\perp})=\left(\frac{1-x}{1-x_0}\right)^{\beta}\Phi_{A(B)}(x_0,\bk_{\perp}).
\end{split}
\end{equation}
where $\beta$ is a parameter. We set $\beta=4$ in our following calculations. The corresponding saturation scale and width of uGD are constants as showed by blue and red dash lines in Fig.~2(a) and 2(b), respectively. 

With above descriptions for gluon production in the full rapidity region, we present in Fig.~4 the two-gluon rapidity correlations as a function of the rapidity gap $\Delta y=y_q-y_p$ for minimum bias samples of $pp$ (red), $AA$ (black) and $pA$ (blue) collisions at 7 TeV. Where the solid, dash and dash-point lines are used to label the results of the small, moderate and large-$x_q$ degree of freedom, respectively. We fix $y_p=0$ in Fig.~4(a) to keep the symmetry of the correlation, and $y_p=-3$ in Fig.~4(b) as a contrast to the case of $y_p=0$. Transverse momentum of two gluon are chosen to make the correlation approximately maximum as demonstrated in the above section. They are $p_{\perp}=q_{\perp}=1.5$  GeV for $pp$ collisions, $p_{\perp}=q_{\perp}=2.5$ GeV for $AA$ collisions, and $p_{\perp}=1.5$ GeV, $q_{\perp}=2.5$ GeV for $pA$ collisions, as indicated in the legend of the figure. 
	 
\begin{figure}[htb]
		\includegraphics[width=\linewidth]{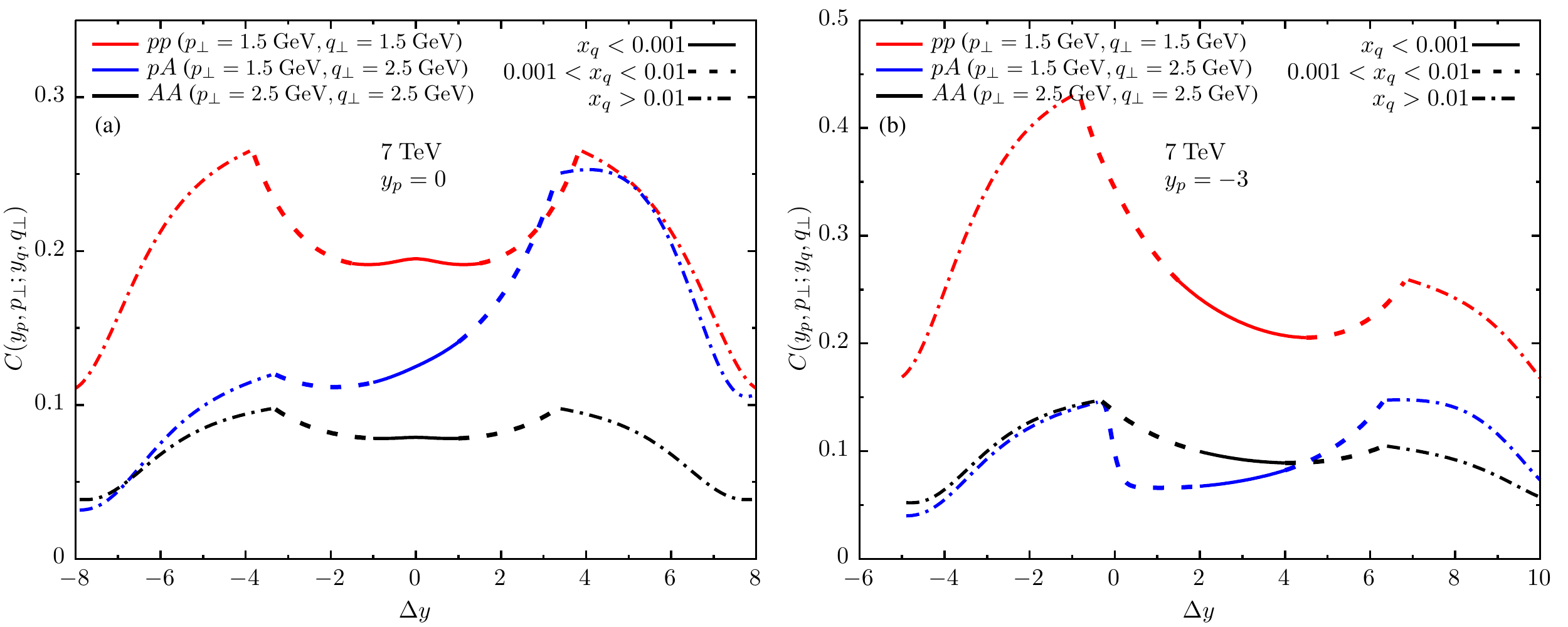}
		\caption{Two-gluon rapidity correlations for minimum bias samples of $pp$ (red), $AA$ (black) and $pA$ (blue) collisions at 7 TeV. Where $\Delta\phi=\phi_q-\phi_p=0$, and the solid, dash and dash-point lines are $x_q<10^{-3}$, $10^{-3}<x_q<10^{-2}$ and $x_q>10^{-2}$, respectively.  $p_{\perp}=q_{\perp}=1.5$ GeV for $pp$ collisions, $p_{\perp}=q_{\perp}=2.5$ GeV for $AA$ collisions, and $p_{\perp}=1.5$ GeV, and $q_{\perp}=2.5$ for $pA$ collisions. (a) $y_p=0$, and (b) $y_p=-3$.}
		\label{fig:rap}
\end{figure}

Fig.~4(a) shows that the correlation pattern in symmetric $pp$ and $AA$ collisions are also symmetry. At the central region, the correlations are almost flat as showed by red and black solid lines. Where the small-$x$ ($x_p,x_q<10^{-3}$) degree of freedom is dominated. The radiated gluon is well evolved by rcBK equation, which is approximately longitudinal boost invariant. So the correlation is almost flat.   

At middle rapidity region as showed by red and black dash lines, the correlations in both $pp$ and $AA$ collisions start to increase. Where the longitudinal momentum fraction of one gluon is larger than $10^{-3}$ and smaller than $10^{-2}$. The correlation is between gluon with moderate and small-$x$ degree of freedom. For the gluon with moderated $x$-degree of freedom, the influence of colour source field has to be taken into account, and the effects of quantum evolution become unimportant. So the correlation between radiated gluon and gluon of colour source field is stronger
than that between two radiated gluon.

At large rapidity region as showed by red and black dash-point lines, the ridges are completely formed in both $pp$ and $AA$ collisions. In this region, the longitudinal momentum fraction of one gluon is larger than $10^{-2}$, and the gluon of colour source field becomes dominate.

In the formalism of CGC, the large-$x$ ($x>10^{-2}$) degrees of freedom act in fact as sources of the small-$x$ degrees of freedom. The evolution towards small-$x$ induces correlations with the color sources. The gluonic color sources are correlated over long distances to ensure that color neutrality. So the ridge indicates the stronger correlation between the gluon of colour source and radiated gluon. 


The formation of the ridge in $pp$ and $AA$ collisions is the same origin. The only difference is there are many nucleons in a nuclei, i.e., a larger saturation scale $Q_s$ and a smaller coupling constant $\alpha_s$. Therefore, the strength of correlations in $AA$ collisions are smaller than those in $pp$ collisions as showed in Fig.~4(a). So the ridge-like correlation pattern is not specified to $AA$ collisions, and can be even easier observed in $pp$ collisions at the same incident energy.

If we change the rapidity position of fixed gluon to a non-zero value, e.g., $y_p=-3$ as showed in Fig.~4(b), the correlation pattern for symmetry collisions will not be symmetry any more. The correlation pattern is dependent on the $x$-component of two selected gluon. If they are both gluon of colour source with large-$x$ degree of freedom, the correlation between them should be the strongest. This is why the maximum correlation in Fig.~4(b) is much stronger than those in Fig.~4(a). 

To asymmetric $pA$ collisions as showed by blue lines in Fig.~4(a), at the central rapidity region, the correlations are not flat due to the fact of gluon produced are asymmetric. The strength of the correlations is between that of $pp$ and $AA$ collisions. It keeps increasing from the negative side to the positive one, in contrast to those in symmetry $pp$ and $AA$ collisions. This is understandable as the nucleus moves along negative direction of beam axis, where the correlation is
weaker than the positive side of the proton. 

From middle to large rapidity region, a small bump and a huge one are formed respectively in the negative and positive sides. This is due to the appearance of the moderate and large-$x$ degree of freedom. So the origin of the bump-like correlations in asymmetric $pA$ collisions is two aspects. One is asymmetry of colliding particles, and other one is the same as that of ridge.

To see how the correlation pattern changes with transverse momentum of trigger gluon and incident energy, we present in Fig.~5 the two-gluon rapidity correlations as a function of the rapidity gap $\Delta y=y_q-y_p$ for three sets of transverse momentum for minimum bias samples of $pp$ collisions at $\sqrt s=7$ TeV (a) and $\sqrt s=0.2$ TeV (b). Where the three sets of transverse momentum of trigger gluon are $p_{\perp}=q_{\perp}=0.8$ GeV (red), $p_{\perp}=q_{\perp}=1.5$ GeV (blue), and $p_{\perp}=q_{\perp}=2.0$ GeV (black) for Fig.~5(a), and $p_{\perp}=q_{\perp}=0.3$ GeV (red), $p_{\perp}=q_{\perp}=0.8$ GeV (blue), and $p_{\perp}=q_{\perp}=2.0$ GeV (black) for Fig.~5(b), respectively.

\begin{figure}[htb]
		\includegraphics[width=\linewidth]{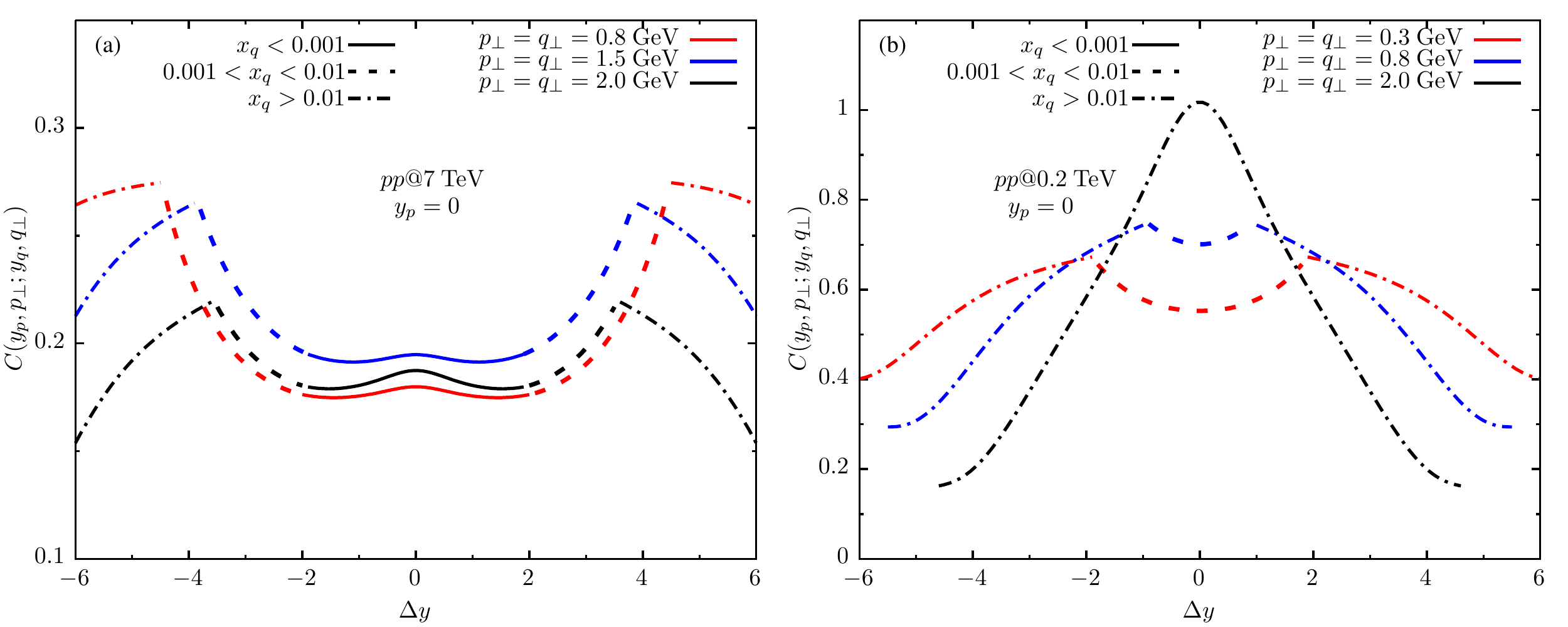}
		\caption{Two-gluon rapidity correlations for minimum bias samples of $pp$ collisions.  Where, $\Delta\phi=\phi_q-\phi_p=0$, $y_p=0$. The solid, dash and dash-point lines are $x_q<10^{-3}$, $10^{-3}<x_q<10^{-2}$ and $x_q>10^{-2}$, respectively. (a) At $\sqrt s=7$ TeV with three sets of transverse momentum of trigger gluon, $p_{\perp}=q_{\perp}=0.8$ GeV (red), $p_{\perp}=q_{\perp}=1.5$ GeV (blue), and $p_{\perp}=q_{\perp}=2.0$ GeV (black). (b) At $\sqrt s=0.2$ TeV with three sets of transverse momentum of trigger gluon, $p_{\perp}=q_{\perp}=0.3$ GeV (red), $p_{\perp}=q_{\perp}=0.8$ GeV (blue), and $p_{\perp}=q_{\perp}=2.0$ GeV (black). }
		\label{fig:rap-en}
\end{figure}

Fig.~5(a) shows again that the highest correlation platform corresponds to $p_{\perp}=q_{\perp}=1.5$ GeV, the solid blue line, as indicated in last section. With the increase of transverse momentum of trigger gluon, the platform, i.e., the solid line, becomes short, and the position of the peak of the ridge moves toward the center, and the height of the ridge becomes lower and lower, as showed by black lines. 

Those features are understandable. From Eq.~(\ref{eq:rapidity in cms}), at given incident energy $\sqrt s$, with increase of transverse momentum, the small-$x$ ($x_q<10^{-3}$) region becomes narrow. At the same condition, from Eq.(\ref{eq:relations between Y and y}), the rapidity of gluon of colour source decreases, and so the saturated momentum. Therefore, the peak position of the ridge moves toward inside, as showed in Fig.~5(a) by the black dash line. 

At lower incident energy, as shown in Fig.~5(b), when transverse momentum is very small, i.e., $p_{\perp}=q_{\perp}=0.3$ GeV and $p_{\perp}=q_{\perp}=0.8$ GeV, there is no solid but only dash lines exist in central rapidity region. In the case, the ridge is still visible. With increase of transverse momentum, the dash line disappears, only dash-point line leaves in the central rapidity region, and the ridge merges to a peak. So the condition of the ridge appearance is a small-$x$ degree of freedom in the central rapidity region. If no small-$x$ degree of freedom, there will be only the strongest correlation between gluon of colour source, i.e., one peak in central rapidity
region. 
  
\section{Summary}
In the framework of CGC, we calculate the two-gluon rapidity correlations of strong colour fields for minimum bias samples of $pp$, $pA$ and $AA$ collisions, respectively. 

To properly choose the transverse momentum of correlated gluon, we first present transverse momentum dependence of the two-gluon rapidity correlation. It shows that at given rapidity gap, the correlation has a peak. In the central rapidity region, the maximum correlation appears when the transverse momentum of two triggered gluon are both close to the sum of saturated momentum of two colliding particles. 

The correlations of $pp$ collisions are distributed in a narrower transverse momentum region than that of $AA$ collisions. This is due to the narrower distribution of uGD of proton. 
The correlations of $pp$ collisions are much stronger than that in $AA$ collisions. This difference simply comes from the saturated momentum scale of colliding particles.

For small rapidity gap, the correlation is insensitive to the width of the rapidity gap in both $pp$ and $AA$ collisions. However, it becomes sensitive in $pA$ collisions due to the different correlation strength caused by the colliding proton and nucleus, which make the correlation change constantly with rapidity gap.

Therefore, the features of the transverse momentum dependence of two-gluon rapidity correlations 
are closely related to the saturation momentum and width of uGD of colliding particles.  

Then we show various $x$ degrees of freedom in the two-gluon rapidity correlation as the function of the rapidity gap. If one triggered gluon is fixed at central rapidity $y_p=0$, ridge-like rapidity correlations are obtained in both symmetry $pp$ and $AA$ collisions, and the huge bump-like rapidity correlations appear in asymmetry $pA$ collisions. It is demonstrated that the correlation platform in the central rapidity region is caused by the correlation between two gluon with the small-$x$ degree of freedom, which is fully evolved by approximately boost invariant rcBK equation. In the middle and large rapidity region, the ridge is gradually formed due to the stronger correlations between gluon with the small and moderate to large-$x$ degree of freedom, where the latter could be the gluon of colour source, which becomes dominated in large rapidity region. 

If we shift the rapidity of the fixed gluon to non-zero value, the correlation pattern for these symmetry collisions will not be symmetry any more. It depends on the $x$-component of two selected gluon. If they are both gluon of colour source with large-$x$ degree of freedom, the correlation will be the strongest. 

The origin of bump-like correlation in $pA$ collisions is two folds. One relates to the appearance of the ridge, and the other one is asymmetry colliding particles. For given colliding particles, the two-gluon rapidity correlation pattern is determined by their $x$ degree of freedoms.   

It is also shown that the ridge is more likely observed at higher incident energy and lower transverse momentum of trigger gluon if the small-$x$ degree of freedom exists in the central rapidity region.
  
Those features of two-gluon rapidity correlation determined in the initial state may be propagated to the final state, and may be observable in RHIC and LHC experimental data. So it is interesting to examine those features by experimental data.

\section*{Acknowledgments} 
We are very grateful for valuable comments of Dr. Bowen Xiao, Raju Venugopalan, and Kevin Dusling. This work is supported in part by the Major State Basic Research Development Program of China under Grant No. 2014CB845402, the NSFC of China under Grant No. 11221504 and the MOE of China under Grant No. QLPL201416. 
	
\bibliographystyle{elsarticle-num}


\end{document}